\documentclass[    final            % use final for the camera ready runs
  ]
  {aipproc}

\layoutstyle{6x9}

\newcommand{\as}{\alpha_s}
\newcommand{\asmz}{\alpha_s(M_Z)}
\newcommand{\aspT}{\alpha_s(p_T)}
\newcommand{\asmur}{\alpha_s(\mu_r)}

\newcommand{\fastnlo}{{\sc fastnlo}}
\newcommand{\nlojet}{{\sc nlojet++}}

\newcommand{\ppbar}{p{\bar{p}}}

\newcommand{\Rtt}{R_{\mbox{3/2}}}
\newcommand{\ptmax}{p_{T\mbox{\footnotesize max}}}
\newcommand{\ptmin}{p_{T\mbox{\footnotesize min}}}

\begin{document}
\hspace{3.5in} \mbox{FERMILAB-CONF-11-285-E}
\vskip-5mm

\title{Determination of the Strong Coupling Constant 
 and Multijet Cross Section Ratio Measurements}

\classification{13.87.-a, 12.38.Qk, 13.87.Ce}
\keywords{Jet production, Quantum Chromodynamics, Strong coupling constant}

\author{M. Wobisch}{
  address={\vskip-1mm(for the D\O\  Collaboration) \\
Louisiana Tech University, Ruston, Louisiana 71272, USA},
}

\begin{abstract}
Concepts and results of determinations of the strong coupling 
constant in hadron collisions are discussed.
A recent $\as$ result from the inclusive jet cross section
in $\ppbar$ collisions at $\sqrt{s}=1.96\,$TeV is presented
which is based on perturbative QCD calculations beyond next-to-leading order.
Emphasis is put on the consistency of the conceptual approach.
Conceptual limitations in the approach of extracting $\as$ 
from cross section data are discussed and how these can be 
avoided by using observables that are defined as ratios of cross sections.
For one such observable, the multijet cross section ratio $\Rtt$,
preliminary results are presented.
\end{abstract}

\maketitle

% ******************************************************************
\section{Introduction}

The strong coupling constant, $\as$, is one of the fundamental 
parameters of the Standard Model of Particle Physics.
The energy dependence of $\as$ is predicted by the renormalization group 
equation (RGE).
The value of $\as$ has been determined in many different processes,
including a large number of results from hadronic jet production,
in either $e^+e^-$ annihilation or in deep-inelastic $ep$ scattering 
(DIS) up to energies of $209\,$GeV~\cite{Bethke:2009jm}.
Prior to the analysis presented in this article, however, 
only a single result had been obtained
from jet production in hadron-hadron collisions.
This $\as$ result is
$\asmz = 0.1178 
        ^{+0.0081}_{-0.0095} (\mbox{exp.}) 
        ^{+0.0071}_{-0.0047} (\mbox{scale}) 
        \pm 0.0059 (\mbox{PDF})    $,
extracted by the CDF collaboration from the inclusive jet cross section 
in $\ppbar$ collisions at $\sqrt{s}=1.8\,$TeV~\cite{Affolder:2001hn}.
All individual uncertainty contributions for this result are larger
than those from comparable results from $e^+e^-$ annihilation or 
DIS~\cite{Bethke:2009jm}.

The first part of this article presents a recent $\as$ determination
from the D\O\ collaboration 
from the inclusive jet cross section
with significantly improved precision.
Conceptual limitations of the approach are discussed, 
and how those are addressed in the D\O\ analysis.
The second part introduces a new observable
to which these conceptual limitations do not apply, 
and which will be valuable for future $\as$ determinations
in new energy regimes accessible at the Tevatron and at the LHC.

% ****************************************************************
\section{Determination of the Strong Coupling Constant}

A new D\O\ analysis~\cite{Abazov:2009nc} extracts the value of $\as$
from inclusive jet cross section data
in $\ppbar$ collisions at $\sqrt{s}=1.96\,$TeV.
It is based on a recent D\O\ measurement of the inclusive jet cross 
section~\cite{:2008hua} 
with unprecedented precision at a hadron collider.
The perturbative QCD (pQCD) prediction for the inclusive jet 
cross section is given by
\begin{equation}
\sigma_{\mbox{pert}}(\as) = 
     \left(  \sum_n \as^n c_n \right) \otimes f_1(\as) \otimes f_2(\as)
     \, ,
\label{eq:pQCD}
\end{equation}
where the $c_n$ are the perturbative coefficients,
the $f_{1,2}$ are the parton distribution functions (PDFs)
of the initial state hadrons,
and the ``$\otimes$'' sign denotes the convolution over 
the momentum fractions $x_1$, $x_2$ of the hadrons.
The sum runs over all powers $n$ of $\as$ which contribute to the 
calculation.
The D\O\ result is based on NLO pQCD ($n=2,3$) plus 
2-loop contributions from threshold corrections~\cite{Kidonakis:2000gi}
($n=4$). 
The latter reduce the scale dependence of the calculations,
leading to a significant reduction of the corresponding uncertainties.
While the $f_{1,2}$ have no explicit $\as$ dependence, our knowledge
of $f_{1,2}$ depends on $\as$ (due to $\as$ assumptions in the PDF analyses).
Since the RGE uniquely relates the value of $\asmur$ at any scale $\mu_r$
to the value of $\asmz$, all equations can be expressed in terms of $\asmz$. 
The total theory prediction for inclusive jet production is given by the pQCD 
result in (\ref{eq:pQCD}), multiplied by a correction factor for 
non-perturbative effects
\begin{equation}
 \sigma_{\mbox{theory}}(\asmz) =  
   \sigma_{\mbox{pert}}(\asmz)
    \cdot  c_{\mbox{non-pert}} \, .
\label{eq:allQCD}
\end{equation}
The pQCD results are computed in \fastnlo~\cite{Kluge:2006xs}
which is based on \nlojet~\cite{Nagy:2003tz,Nagy:2001fj}
and the calculations from Ref.~\cite{Kidonakis:2000gi}.
To determine $\asmz$, recent PDF results are used
and $\asmz$ is varied in $\sigma_{\mbox{pert}}(\asmz)$
(i.e.\ simultaneously in the matrix elements and in the PDFs)
until $\sigma_{\mbox{theory}}(\asmz)$ agrees with the data.
There are, however, two conceptual issues when extracting $\as$ from
cross section data.
\begin{enumerate}

\item
When performing the DGLAP evolution of the PDFs,
all PDF analyses are assuming the validity of the RGE
which has so far been tested only for energies up to 209\,GeV.
Since extracting $\as$ at higher energies means testing 
(and therefore questioning) the RGE,
using these PDFs as input would be inconsistent.

\item
D\O\ jet data have been used in all recent global PDF analyses.
The PDF uncertainties are therefore correlated with the experimental 
uncertainties in those kinematic regions in which 
the D\O\ jet data had strong impact on the PDF results.
As shown in Figs.~51--53 in Ref.~\cite{Martin:2009bu}, 
this is the case for the proton's  gluon density
at $x > 0.2 - 0.3$.
Since the correlations between PDF uncertainties and 
experimental uncertainties are not documented, 
the $\as$ extraction should avoid
using those data points which already
had significant impact on the PDF results.

\end{enumerate}
In light of the second issue, the D\O\ $\as$ extraction
uses only data points which are insensitive to $x > 0.2 - 0.3$.
Since all of these data points have $p_T$ below 145\,GeV,
the first issue does not become relevant here.
This leaves 22 (out of 110) inclusive jet data points.

\begin{figure}
  \includegraphics[width=8.4cm]{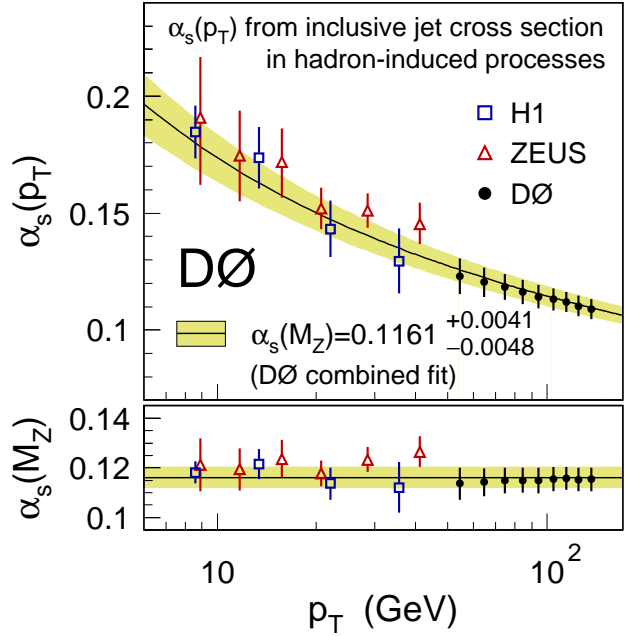}
  \caption{\label{fig:fig1}
   Recent D\O\ results from a determination
   of the strong coupling constant from inclusive jet cross section data,
   compared to corresponding results in DIS fom HERA.
  }
\end{figure}

The $\as$ extraction
uses PDFs from the MSTW2008 analysis~\cite{Martin:2009iq}
which were obtained at NNLO
(consistent with the precision of the 
theory calculation used here).
These PDFs have been determined for 21 $\asmz$ values between
$0.107$ and $0.127$~\cite{Martin:2009bu}.
The continuous $\asmz$ dependence of the pQCD cross sections is
obtained by interpolating the cross section results
for the PDF sets for different $\asmz$ values.
PDF uncertainties are computed using the twenty uncertainty 
eigenvectors (corresponding to 68\%~C.L.).
The uncertainties in the pQCD calculation due to uncalculated higher-order 
contributions are estimated from the $\mu_{r,f}$ dependence of the 
calculations when varying the renormalization and factorization 
scales in the range $0.5 \le (\mu_{r,f}/p_T) \le 2$.
In a first step, data points with same  $p_T$ are combined to determine
nine values of $\aspT$ for $50 < p_T < 145\,$GeV.
These results are shown in Fig.~\ref{fig:fig1} 
and compared to results obtained in DIS.
A combined determination from all 22 data points yields a result of
\begin{equation}
\asmz \, = \, 0.1161   \,
 ^{+0.0034}_{-0.0033} \, \mbox{(exp.)}    \;
 ^{+0.0010}_{-0.0016} \, \mbox{(non-pert.)}   \;
 ^{+0.0011}_{-0.0012} \, \mbox{(PDFs)}   \;
^{+0.0025}_{-0.0029} \, \mbox{(scale)}   \, .
\end{equation}
This is currently the most precise result from a hadron collider,
with similar precision as recent results from jet production in DIS.

% ****************************************************************
\section{Measurement of Multijet Cross Section Ratios}

\begin{figure}
  \includegraphics[height=6cm]{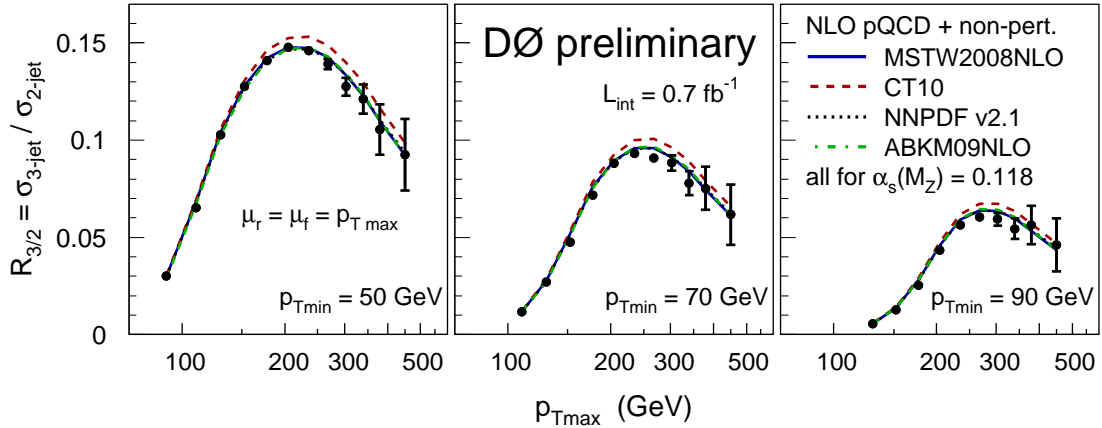}
  \caption{\label{fig:fig2} 
   Preliminary D\O\ results for the multijet cross section ratio $\Rtt$,
   measured as a function of $\ptmax$ for different $\ptmin$ requirements.
   Theory calculations for $\asmz = 0.118$ and for different PDFs
   are compared to the data.
  }
\end{figure}

The conceptual issues discussed above, which limit the $x$ range, 
and therefore also the $p_T$ range of the data points used in 
the $\as$ determination, are both related to the fact that 
the observable is sensitive to the proton PDFs 
which are required as external input in the $\as$ determination.
These limitations can be avoided by studying observables
which are largely independent of the PDFs, but still sensitive to $\as$.
One class of such observables are ratios of multijet cross sections.
The variable $\Rtt$ represents the conditional probability 
that a given inclusive dijet event also has a third jet.
It is defined as the ratio of the inclusive 3-jet and 
dijet cross sections and investigated as a function of $\ptmax$,
the transverse momentum of the leading jet in an event, which
is a common scale for the 3-jet and the dijet production processes.
Therefore $\Rtt(\ptmax)$ is directly sensitive to $\as$ 
at the scale $\mu_r=\ptmax$ while the PDFs cancel to a large extent 
in the cross section ratio.
Technically, the $n$-jet cross section (for $n=2,3$) is defined 
by all events with $n$ or more jets with $p_T$ above $\ptmin$,
in a given rapidity region (here: $|y|<2.4$ for the $n$ leading jets).
The preliminary results of a recent D\O\ measurement of 
$\Rtt$~\cite{D0:2010rr}, obtained for different values
of $\ptmin = 50,\, 70,\, 90\,$GeV, are corrected to particle level
and presented in Fig.~\ref{fig:fig2} as a function of $\ptmax$.
The data are well described by theory calculations
based on NLO pQCD plus non-perturbative corrections,
for different PDFs~\cite{Martin:2009bu,Lai:2010vv,Ball:2011mu,Alekhin:2009ni}
using in all cases $\asmz=0.118$ (in the matrix elements and in the PDFs).
In the future, measurements of $\Rtt$ and related observables will provide 
a solid basis for determinations of $\as$ over the whole $p_T$ region, 
accessible at the Tevatron and the LHC, 
and without conceptual issues related to the proton PDFs.
Such results will allow testing the RGE in a novel energy regime.

% ****************************************************************

\end{document}